\documentclass[aps,prd,nopacs,floatfix,notitlepage,superscriptaddress,nofootinbib,twocolumn,a4paper,longbibliography]{revtex4-1}
\usepackage{amsfonts,amsmath,units,wasysym,epsfig,graphicx,verbatim,color,subfigure,graphicx,bm,mathrsfs,lipsum,hyperref,cleveref}
\usepackage{booktabs}
\usepackage[normalem]{ulem}  

\begin{document}

\newcommand{\bhaskar}[1]{\textcolor{red}{\bf BB: #1}}
\newcommand{\pc}[1]{\textsf{\color{blue}{\bf #1}}}

\newcommand{\USAL}{Departamento de F\'isica Fundamental and IUFFyM, Universidad de Salamanca, Plaza de la Merced S/N, E-37008 Salamanca, Spain}

\newcommand{\Uliege}{Space Sciences, Technologies and Astrophysics Research (STAR) Institute, Universit\'e de Li\`ege, B\^at. B5a, 4000 Li\`ege, Belgium}

\newcommand{\HU}{Hamburger Sternwarte, Gojenbergsweg 112, D-21029 Hamburg, Germany}

\title{Compact object of HESS J1731-347 and its implication on neutron star matter}

\author{Prasanta Char}
\email{prasanta.char@usal.es}
\affiliation{\USAL}\affiliation{\Uliege}

\author{Bhaskar Biswas}
\email{phybhaskar95@gmail.com}
\affiliation{\HU}


\begin{abstract}
In this work, we investigate the impact of the possibility of a small, subsolar mass compact star, such as the recently reported central compact object of HESS J1731-347, on the equation of state (EOS) of neutron stars. We have used a hybrid approach to the nuclear EOS developed recently where the matter around nuclear saturation density is described by a parametric expansion in terms of nuclear empirical parameters and represented in an agnostic way at higher density using piecewise polytropes. We have incorporated the inputs provided by the latest neutron skin measurement experiments from PREX-II and CREX, simultaneous mass-radius measurements of pulsars PSR J0030+0451 and PSR J0740+6620, and the gravitational wave events GW170817 and GW190425. The main results of the study show the effect of HESS J1731-347 on the nuclear parameters and neutron star observables. Our analysis yields the slope of symmetry energy $L=45.71^{+38.18}_{-22.11}$ MeV, the radius of a $1.4 M_\odot$ star, $R_{1.4}=12.18^{+0.71}_{-0.88}$ km, and the maximum mass of a static star, $M_{\rm max}= 2.14^{+0.26}_{-0.17} M_\odot$ within $90\%$ confidence interval, respectively.
\end{abstract}

\maketitle

\section{Introduction}
Recent observation of a compact object inside the supernova remnant HESS J1731-347 has provided an estimation of its mass and radius to be $M = 0.77^{+0.20}_{-0.17} M_\odot$ and $R = 10.4^{+0.86}_{-0.78}$ km at $1\sigma$ credible levels, respectively \cite{Doroshenko:2022na}. Understandably, this observation has generated a lot of attention from the nuclear physics community working on the dense matter physics beyond the normal nuclear matter density \cite{Brodie:2023pjw,Huang:2023vhy,Li:2023vso, Kubis:2023gxa, Koehn:2024set, Miao:2024vka}. Different possibilities have been explored to characterize this object, such as, a strange quark or a hybrid star \cite{DiClemente:2022wqp,Horvath:2023uwl,Oikonomou:2023otn,Rather:2023tly,Yuan:2023dxl,Laskos-Patkos:2023tlr,Gao:2024chh,Mariani:2024gqi}, or a dark matter admixed compact star \cite{Sagun:2023rzp,Routaray:2023txs}.
A typical neutron star (NS) usually has a mass within the range of $\sim 1-2 M_\odot$. NSs being extremely dense compact objects have profound effects on our understanding of dense nuclear matter higher than nuclear saturation density ($\rho_0$) \cite{Glendenning:1997wn}. The equation of state (EOS) of highly dense matter is connected to the observations of the masses and radii of the compact objects through the solutions of the Tolman-Oppenheimer-Volkoff equations. Hence, each new source with observed mass and radius allows us to test and constrain our models of nuclear physics and increase precision over the previous results. Therefore, the compact object of HESS J1731-347 being an outlier in the distribution of observed pulsar masses may provide a unique insight to neutron star matter. However, one must take this estimation with caution as other studies have indicated several assumptions on the atmospheric composition and distance that may have led to these extreme numbers \cite{Alford:2023waw}. In particular, \textcite{Alford:2023waw} have pointed out that the results of \textcite{Doroshenko:2022na} crucially depend on the following assumptions: i) the object has uniform-temperature carbon atmosphere, ii) the distance of this object is 2.5 kpc. If the distance is considered to be 3.2 kpc, the inferred mass would be much higher $\sim 1.4 M_\odot$. They also have shown that the XMM-Newton spectra of this object is a poor fit with the assumed atmosphere model. As a consequence, further analyses are required to examine validity of the assumptions to fit the data. It is also important to mention that some latest supernova simulations are not able to find a remnant less than $1.19 M_\odot$ \cite{Muller:2024aod}. Even though the results remain controversial, there has been a string of theoretical studies that investigates the nature of that object and its implications. For example, \textcite{Horvath:2023uwl} have performed a minimal consistency check for the compact object of HESS J1731-347 within the existing theoretical models of compact stars. \textcite{Zhang:2024ldq} have proposed a formation scenario for low mass neutron stars. 

In any case, this situation provides an opportunity to check the consistency of such extreme observations with our existing models of neutron star matter, if they are really confirmed in the future. 

Our understanding of NS matter and the nuclear equation of state (EOS) have improved drastically after the several astrophysical observations in the previous decades, such as the massive pulsars ($\gtrsim 2 M_\odot$) \cite{Antoniadis:2013pzd,Cromartie:2019kug,Fonseca:2021wxt}, and  the first ever observation of gravitational waves (GW) and the electromagnetic counterparts from the binary neutron star merger event GW170817, reported by the LIGO-Virgo-Kagra (LVK) collaboration \cite{TheLIGOScientific:2017qsa,LIGOScientific:2017ync}. The GW observation has provided important information on the combined tidal deformability of the binary system which is an EOS-dependent quantity. Then, the NICER collaboration has also provided simultaneous mass-radius measurements of PSR J0030+0451 and PSR J0740+6620 \cite{Riley:2019yda,Miller:2019cac,Riley:2021pdl,Miller:2021qha}. From the perspective of ground-based experiments, the measurements of neutron skin thickness of $^{208}$Pb by PREX \cite{PREX:2021umo} and $^{48}$Ca by CREX \cite{CREX:2022kgg} collaborations, respectively, have shed a new light on the nuclear symmetry energy and its slope parameter. In parallel, there have also been advancements in the ab-initio calculations of the properties of pure neutron matter with chiral effective field theories \cite{Hebeler:2013nza,Tews:2012fj,Lynn:2015jua,Drischler:2015eba,Drischler:2017wtt,Huth:2020ozf,Drischler:2021kxf}. These calculations have provided major insights to the properties of nuclear matter around and below nuclear saturation.

In this paper, we investigate the influence of the PREX, and CREX measurements on the EOS, along with the mass-radius data of HESS J1731-347, assuming it to be a neutron star. We introduce these constraints successively and check their effects on the nuclear matter properties. The paper is organized in the following way. In section \ref{EOS}, we briefly describe our EOS model. The relevant constraints used in the work and the Bayesian methodology are discussed in section \ref{method}. Finally, we discuss our findings in section \ref{res} and summarize in section \ref{summary}.

\section{EOS of NS Matter}
\label{EOS}
In this section, we provide a brief overview of our approach to model the supranuclear matter inside the NS. We have used the hybrid nuclear + PP EOS parametrization that has been developed by \textcite{Biswas:2020puz} and used to provide multimessenger analyses of NS properties \cite{Biswas:2020xna,Biswas:2021yge}. It has also been used to create a framework to infer Hubble parameter directly from GW signals from the future binary neutron star BNS mergers \cite{Ghosh:2022muc,Ghosh:2024cwc}. The core EOS consists of two main parts: a) nuclear physics informed expansion-based near saturation, b) agnostic polytropic at high densities. Around $\rho_0$, we have used an EOS based on Taylor's series expansion of the isoscalar and isovector components of the energy per nucleon. 

First the energy per particle is expanded in terms of asymmetry up to second order.
\begin{equation}
    e(\rho,\delta) \approx  e_0(\rho) +  e_{\rm sym}(\rho)\delta^2,
\end{equation}
where, $e_0(\rho)$ is the energy per particle for symmetric nuclear matter (SNM), and $e_{\rm sym}(\rho)$ is the symmetry energy.
At $\rho_0$, they are further expanded in Taylor's series and the coefficients are defined as the nuclear empirical parameters. 

\begin{eqnarray}
 e_0(\rho) &=&  e_0(\rho_0) + \frac{ K_0}{2}\chi^2 \label{eq:e0} +\,...,\\
e_{\rm sym}(\rho) &=&  e_{\rm sym}(\rho_0) + L\chi + \frac{ K_{\rm sym}}{2}\chi^2 
 ..., \label{eq:esym}
\end{eqnarray}
where $\chi \equiv (\rho-\rho_0)/3\rho_0$ expresses the deviation from $\rho_0$. We keep the terms upto second order as the higher order terms do not contribute significantly to the EOS around $\rho_0$. We also fix the saturation density, $\rho_0 = 0.16$ fm$^{-3}$ and binding energy per nucleon of SNM at saturation, $e_0(\rho_0) = -15.9$ MeV \cite{Brown:2013pwa, Margueron:2017eqc,Tsang:2019ymt}.  Therefore the free parameters in this model are the incompressibility ($K_0$), nuclear symmetry energy ($e_{\rm{sym}}$) and its slope parameter ($L$), and the isovector incompressibility ($K_{\rm{sym}}$) at saturation. We consider the previous theoretical and experimental works to estimate these quantities \cite{Oertel:2016bki,Dutra:2012mb, Dutra:2014qga} and take a unifrom prior over a large range of value to incorporate their uncertainties [see Table \ref{tab1}]. For the cold NS matter, the additional conditions of $\beta$-equilibrium and charge neutrality are also imposed. 
Then, at higher densities we have used an agnostic EOS with piecewise polytrope (PP) parametrization \cite{Read:2008iy}. In particular, we have used an arrangement with a three-segment polytrope ($\Gamma_1, \Gamma_2, \Gamma_3$ being the polytropic indices) starting at $1.25 \rho_0$. The next two stitching points are $1.8$ and $3.6 \rho_0$, respectively. Hence, we have a total of seven parameters, $\mathcal{E}=\{K_0, e_{\rm{sym}}, L, K_{\rm{sym}}, \Gamma_1, \Gamma_2, \Gamma_3\}$, representing the EOS of NS matter. For the crust EOS, we have used the standard Baym-Pethick-Sutherland EOS table \cite{Baym:1971pw}.

\begin{figure*}
    \centering
    \includegraphics[width=\textwidth]{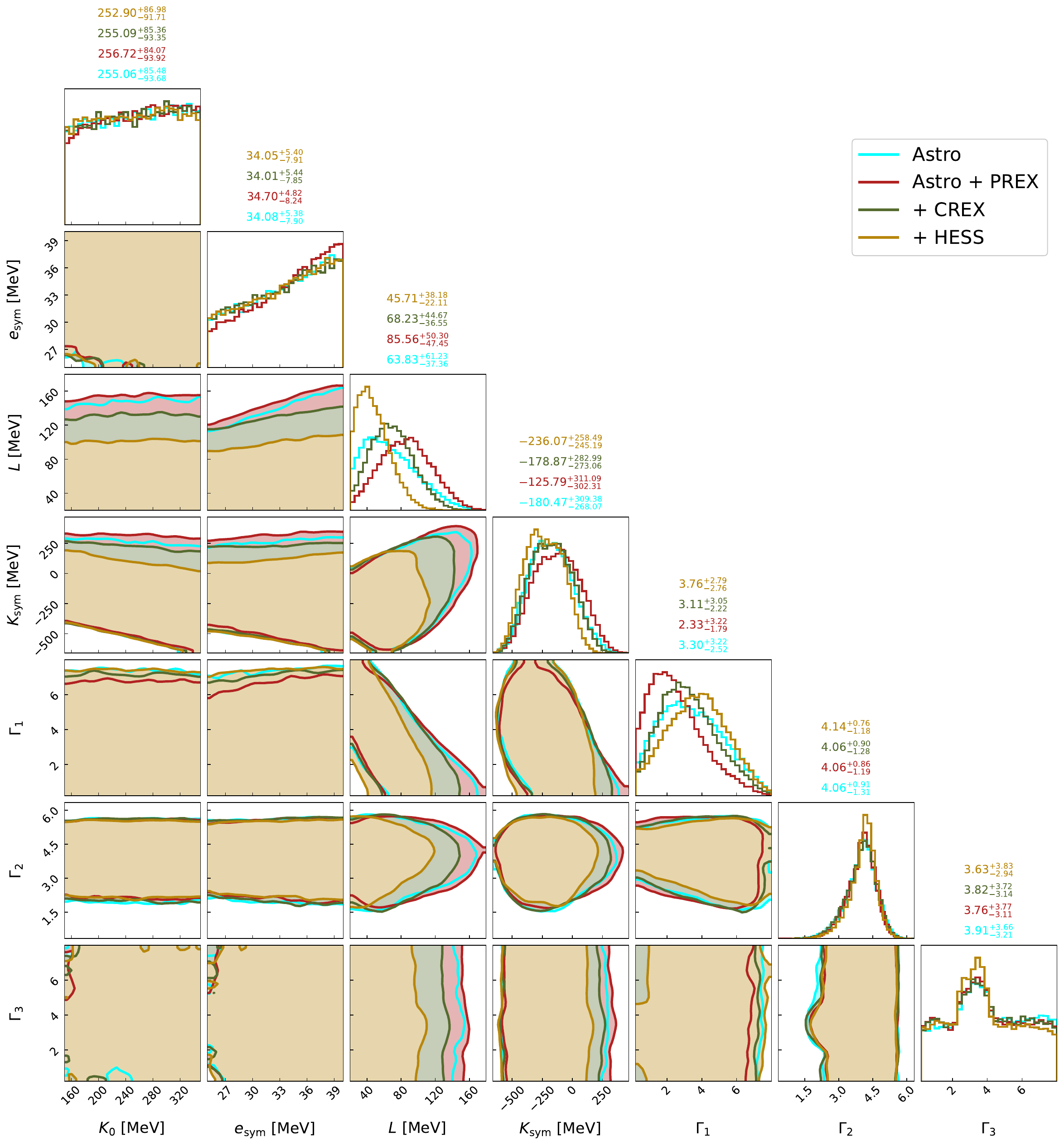}
\caption{Posterior distribution of EOS parameters with their 90\% CI.}
    \label{fig:eos_params}
\end{figure*}

\begin{figure}
    \centering
    \includegraphics[width=0.45\textwidth]{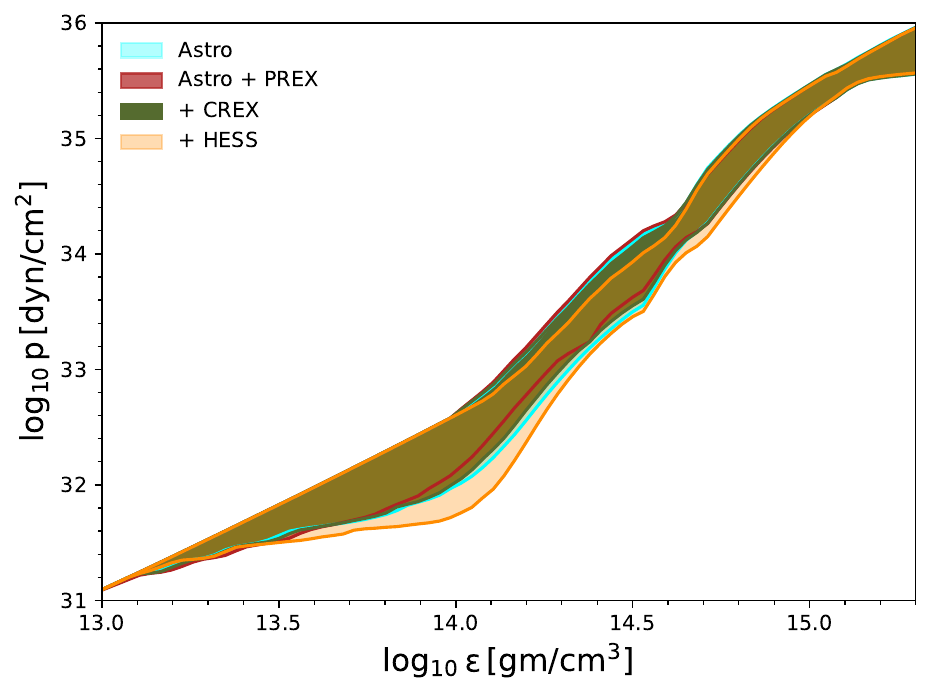}
\caption{Posterior distribution of pressure as a function of energy density at 90 \% CI. The axes are shown in the logarithmic scale.}
    \label{fig:eos-post}
\end{figure}

\begin{figure}
    \centering
    \includegraphics[width=0.45\textwidth]{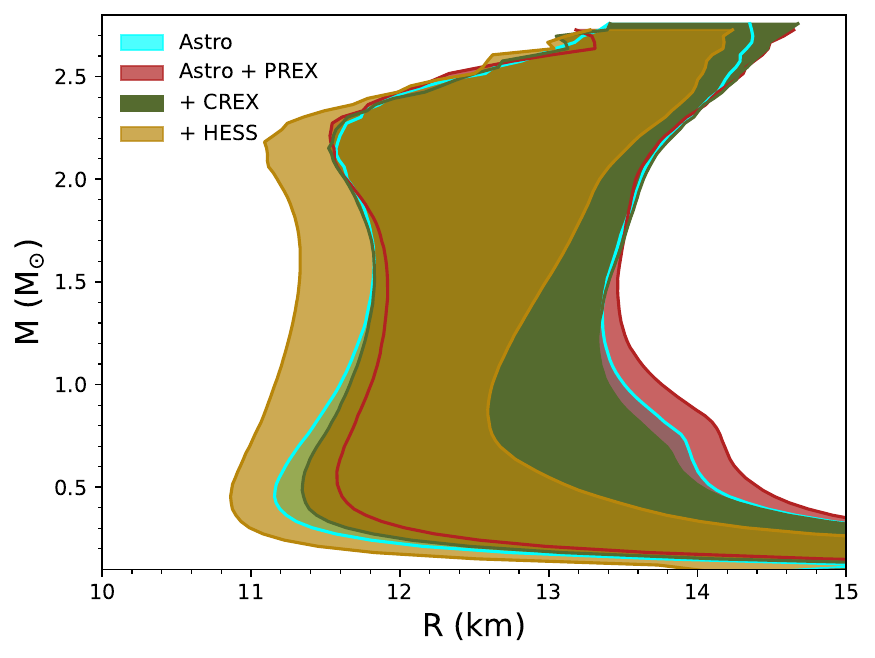}
\caption{Mass-radius posteriors at 90 \% CI with the EOS models shown in figure \ref{fig:eos-post}. }
    \label{fig:mr-post}
\end{figure}

\begin{figure*}
    \centering
    \includegraphics[width=\textwidth]{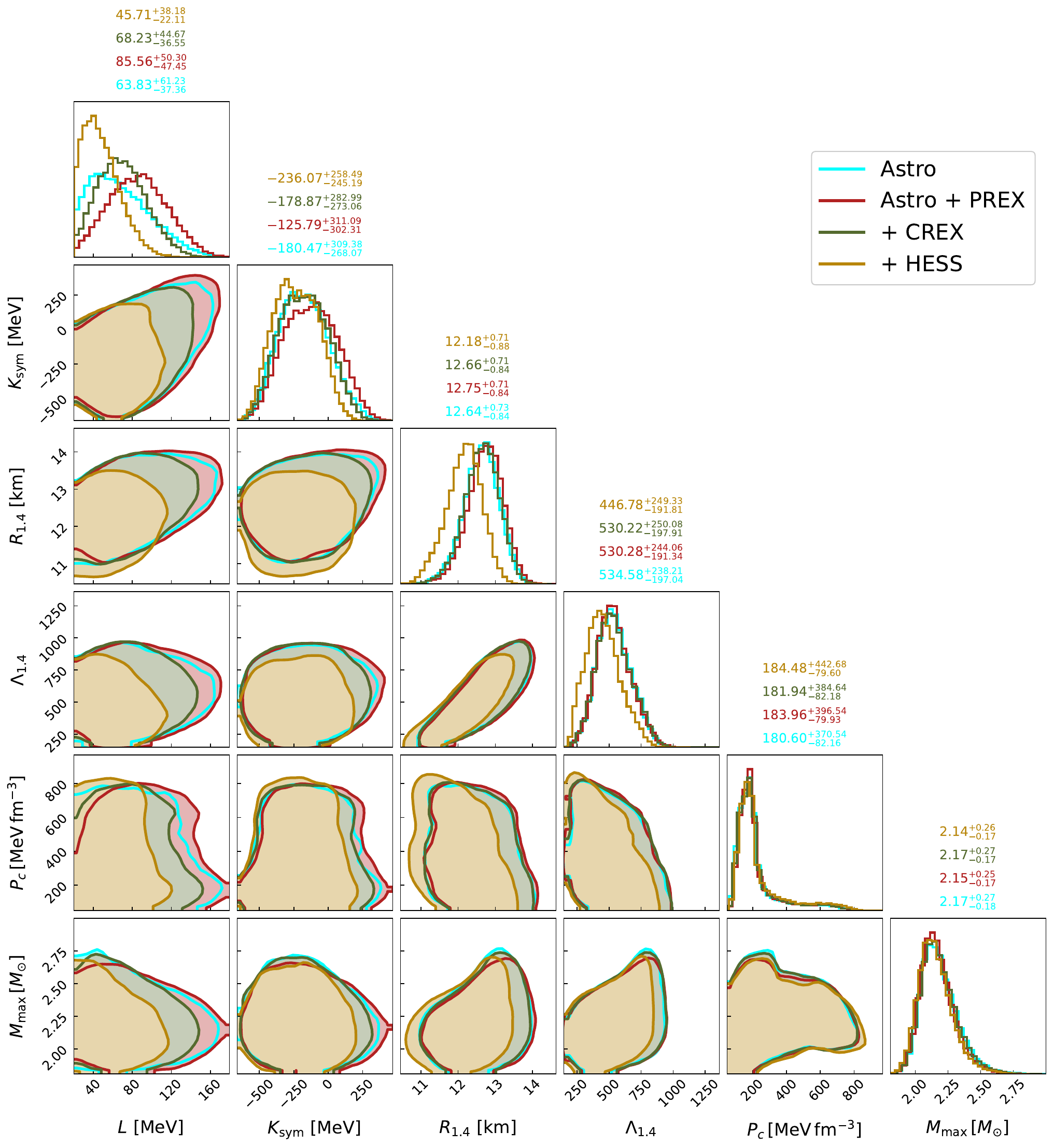}
\caption{Correlation plot for isovector quantities with neutron star observables with their 90\% CI.}
    \label{fig:macro_params}
\end{figure*}

\begin{figure*}
    \centering
    \includegraphics[width=\textwidth]{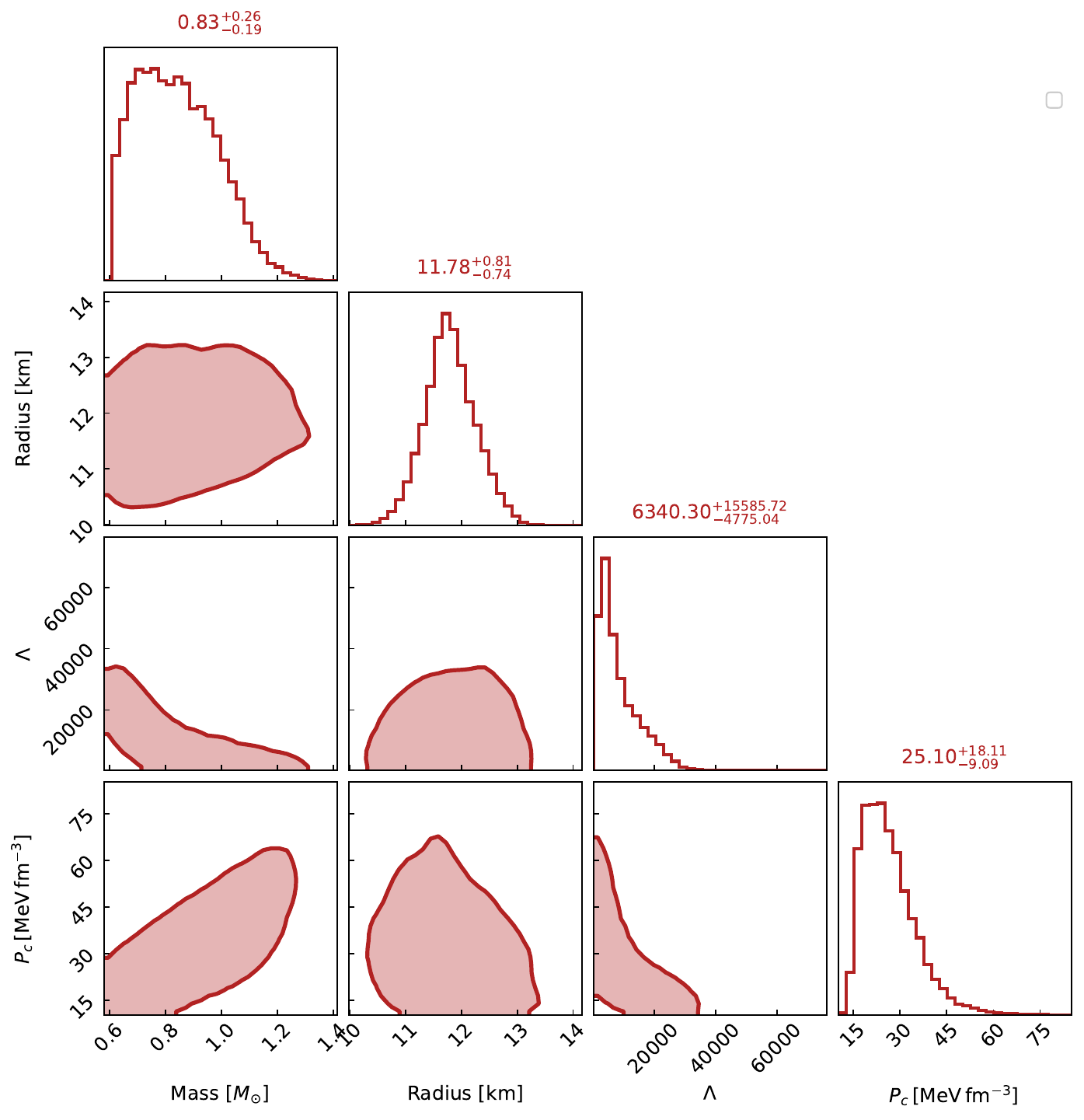}
\caption{Correlation plot of inferred properties of HESS J1731-347 at 90\% CI.}
    \label{fig:HESS_prop}
\end{figure*}

\begin{figure}
    \centering
    \includegraphics[width=0.45\textwidth]{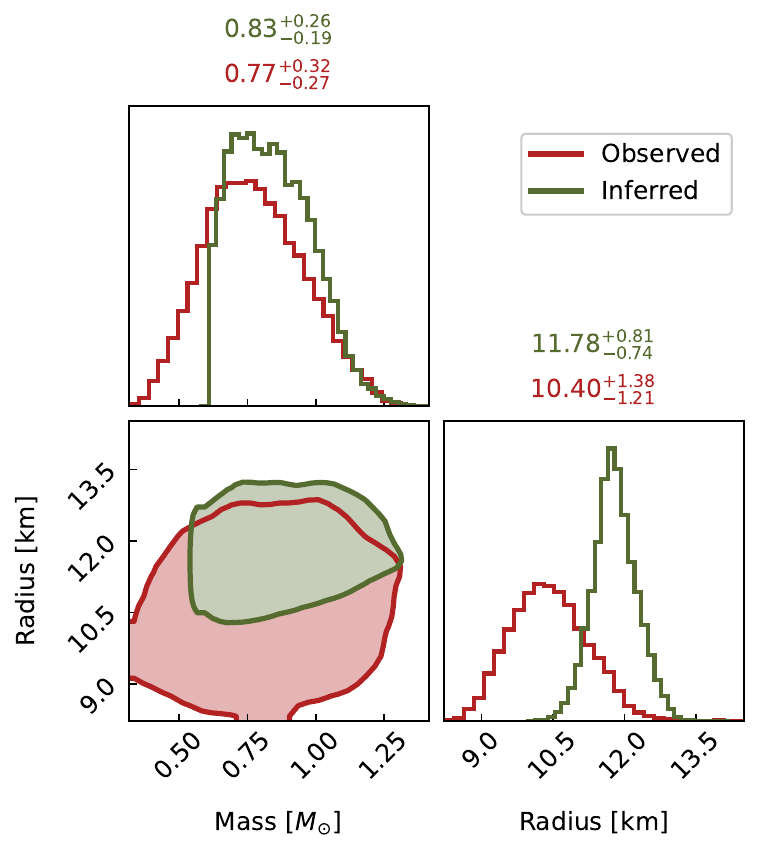}
\caption{Comparison between the actual data and inferred mass-radius posteriors of HESS J1731-347 at their 90\% CI.}
    \label{fig:HESS_MRs}
\end{figure}
\section{Methodology}
\label{method}
In this section, we describe briefly the inference methodology used in this work.  We have used the tidal deformability posteriors from two BNS merger events GW1701817, and GW190425 \cite{advanced-ligo,advanced-virgo,LIGOScientific:2018hze,Abbott:2020uma}. We have used the mass-radius posteriors of PSR J0030+0451 and PSR J0740+6620 \cite{riley19c,zenodo_riley_0740,miller_2019_3473466,zenodo_miller_0740}. For the HESS J1731-347, we have used the mass-radius (M-R) posterior \cite{doroshenko_2022_6702216} assuming the uniformly emitting carbon atmosphere provided by \textcite{Doroshenko:2022na}. In particular, we have used the data file, ``xray\_only\_carbatm.txt" that includes only X-ray data using single temperature carbon atmosphere model and Gaia parallax priors. We have not used the other dataset that includes EOS priors from previous works as that would create an inconsistency with our own EOS model used in this work. The posterior of the EOS parameters can be expressed as,

\begin{equation}
     P(\mathcal{E} | {d}) \propto  \Pi_i P ({d_i} | \mathcal{E}) P(\mathcal{E}) \,,
\end{equation}
where $d = (d_{\rm LVK}, d_{\rm NICER}, d_{\rm HESS}, d_{\rm PREX},d_{\rm CREX})$ is the set of different constraints used. The individual likelihoods can be written as follows:\\ 

\begin{itemize}
    \item GW Observations: the masses $m_1, m_2$ of the two binary components and the corresponding tidal deformabilities $\Lambda_1, \Lambda_2$. In this case,
\begin{align} \label{eq:GW-likelihood}
    P(d_{\rm LVK}|\mathcal{E}) = \int^{m_{\rm max }}_{m_2}dm_1 \int^{m_1}_{m_{\rm min}} dm_2  P(m_1,m_2|\mathcal{E}) \nonumber\\
    \times P(d_{\rm LVK} | m_1, m_2, \Lambda_1(m_1,\mathcal{E}), \Lambda_2(m_2,\mathcal{E})) \,,
\end{align}
where $m_{\rm max }$ is the maximum mass of a NS for a particular set of E0S parameter $\mathcal{E}$, $P(m_1,m_2|\mathcal{E})$ is the prior distribution over the component masses at the source frame. This is determined by the population modelling of NSs.
\begin{align}
 P(m_1,m_2|\mathcal{E}) = P(m_1|\mathcal{E}) \times P(m_2|\mathcal{E})
\end{align}
We take a simple population model define by,
\begin{equation} \label{mass_dist}
    P(m\mid \bm{\mathcal{E}})= 
\begin{cases}
    \frac{1}{ m_{\text{max}}(\bm{\mathcal{E}}) - m_{\text{min}} },& \text{iff}\ \  m_{\text{min}}\leq m \leq m_{\text{max}} \\
    0,              & \text{otherwise\,,}
\end{cases}
\end{equation}
where we have chosen, $m_{\rm min} = 0.1 M_\odot$ in our analyses.
Given the high-precision measurement of the chirp mass in GW observations, equation~\ref{eq:GW-likelihood} can be further simplified by fixing the GW chirp mass to its median value with not so much affecting the result~\cite{Raaijmakers:2019dks} given its high precision measurement. Then we will have one less parameter to integrate over as $m_2$ will be a deterministic function of $m_1$. Hence, we fix it to the observed median value and use it to generate a set of binary neutron star systems and their respective tidal deformability. Then, we evaluate the likelihood directly from the GW data with the help of a Gaussian kernel density estimator (KDE) and used the software package {\tt statsmodels} \cite{seabold2010statsmodels} for this purpose. We have used the publicly available posterior samples of $m_1,m_2,\Lambda_1,\Lambda_2$ distribution of GW170817~\footnote{LVK collaboration,~\href{https://dcc.ligo.org/LIGO-P1800115/public}{https://dcc.ligo.org/LIGO-P1800115/public}} and GW190425\footnote{LVK collaboration,~\href{https://dcc.ligo.org/LIGO-P2000026/public}{https://dcc.ligo.org/LIGO-P2000026/public}}.

   \item X-ray observations: the mass ($m$) and radius ($R$) measurements of NS. Therefore, the corresponding likelihood takes the following form,
\begin{align}
    P(d_{\rm X-ray}|\mathcal{E}) = \int^{m_{\rm max }}_{m_{\rm min }} dm P(m|\mathcal{E})  
    P(d_{\rm X-ray} | m, R(m,\mathcal{E})) \,.
\end{align}
Similar to GW observations, we modelled the likelihood for the X-ray sources with Gaussian KDEs.

   \item Neutron skin thickness measurements: The PREX collaboration reported the measurement of the neutron skin thickness of $^{208}{\rm Pb}$ to be $R_{\rm skin}^{208} = 0.283 \pm 0.071$ fm \cite{PREX:2021umo}. Shortly after, the CREX collaboration reported the neutron skin thickness of $^{48} {\rm Ca}$ to be $R_{\rm skin}^{48} = 0.121 \pm 0.026 ({\rm exp}) \pm 0.024 ({\rm model}) $ fm \cite{CREX:2022kgg}. As in Ref.~\cite{Essick:2021kjb,Biswas:2021yge} , we have used a universal relation  between $r_{\rm skin}$ and empirical parameter $L$ proposed in Ref.~\cite{Vinas:2013hua} for the likelihood computation of PREX-II: 
   \begin{equation}
       R_{\rm skin}^{208} {\rm [fm]} = 0.101 + 0.00147 \times L [\rm MeV].
   \end{equation}
    Similarly for CREX, we also use a similar empirical relation from Ref.~\cite{Tripathy:2020yig}, to evaluate the neutron skin thickness of $^{48} \text{Ca}$ utilizing the value of $R_{\text{skin}}^{208}$.
    \begin{equation}
        R_{\text{skin}}^{48} =  0.0416 +  0.6169 R_{\text{skin}}^{208}.
    \end{equation}
    Finally, we use a Gaussian distribution to model both their likelihood.
\end{itemize}


\begin{table}
 \begin{tabular}{cc}
  \hline 
  Parameter & Prior \\ 
  \hline 
  $K_0$ (MeV) & $\mathcal{U}(150,350)$ \\
  $e_{sym}$ (MeV) & $\mathcal{U}(25,40)$ \\
  $L$ (MeV) & $\mathcal{U}(20,180)$ \\
  $K_{sym}$ (MeV) & $\mathcal{U}(-1000,500)$ \\
  $\Gamma_1$ & $\mathcal{U}(0.2,8)$ \\
  $\Gamma_2$ & $\mathcal{U}(0.2,8)$ \\
  $\Gamma_3$ & $\mathcal{U}(0.2,8)$ \\
  \hline
 \end{tabular}
\caption{Prior distributions of the EOS parameters.}
\label{tab1}
\end{table}

\section{Results}
\label{res}

In this paper, we focus on four different analyses with different combinations of astrophysical constraints and information from the neutron skin measurement experiments. We start with the reanalysis of the EOS inference results obtained in Ref. \cite{Biswas:2021yge} using the GW170817, GW190425, two NICER measurements, and PREX results, but with different priors and look for any effect on the EOS quantities. Then, we have added the information from CREX, and finally the M-R data of the HESS J1731-347. We have sampled the posterior using the nested sampling algorithm implemented in {\tt PyMultiNest} software package \cite{Buchner:2014nha}.

In Figure \ref{fig:eos_params}, we have shown the posterior distributions of the EOS parameters along with their median and 90\% confidence intervals (CI) for different constraints applied. We denote the GW and NICER as ``Astro", then our second case is ``Astro + PREX". The third case has CREX on top of the second case, denoted by ``+ CREX". Finally, the fourth case includes HESS J1731-347 along with the first three, denoted by ``+ HESS". We find that $K_0$ and $e_{\rm sym}$ are not constrained by the astrophysical data used. The distribution of $K_0$ is mostly flat, whereas the we see an increasing trend for $e_{\rm sym}$ to the higher values.  This is a result of choosing a flat prior on these two quantities. In the previous work within the same EOS model \cite{Biswas:2020puz, Biswas:2021yge}, similar trends were seen when uniform priors were used. Also, when Gaussian priors were used, the posterior reflected the priors again confirming that these parameters are insensitive to astrophysical constraints and the PREX measurements. Here, we confirm again that the CREX and HESS measurements also have no effects on $K_0$ and $e_{\rm sym}$. However, the situation is different when we look at $L$ and $K_{\rm sym}$. Taking a Gaussian prior of $\mathcal{N}(58.7,28.1)$ MeV on $L$,  its bound was estimated in Ref. \cite{Biswas:2021yge} to be, $L=54^{+21}_{-20}$ MeV with astrophysical constraints, and $L=69^{+21}_{-19}$ MeV at $1\sigma$ level after adding the PREX-II result. In this case, we have used a large uniform prior on $L$ and found the median value shifted to a higher value, $L=64^{+61}_{-37}$ MeV for Astro, and $L=86^{+50}_{-47}$ MeV when PREX-II is added. Higher $L$ values were reported previously as a consequence of the PREX measurements \cite{Reed:2021nqk,Reed:2023cap}. Then, we see the median of $L$ getting shifted to lower values after successively adding CREX and HESS J1731-347. After applying all the constraints, we find, $L=46^{+38}_{-22}$ MeV. From nuclear experiments, there is not much information available about $K_{\rm sym}$. In the literature, its value is suggested to be negative from most phenomenological interaction calculations \cite{Dutra:2012mb,Dutra:2014qga}. The constraints on $K_{\rm sym}$ from astrophysics is also weak as we found that the 90\% CI is quite large for all cases, but we see the similar trend in the median values, as $L$. With Astro constraints only, the median of $K_{\rm sym}$ becomes $-180$ MeV.  With Astro+PREX, we find a sharp increase in $K_{\rm sym}$ when it reaches $-126$ MeV. But it shifts to lower values of $-179$ MeV again after adding CREX and further down to be $-236$ MeV after adding HESS. We also notice a correlation between $L$ and $K_{\rm sym}$.  Similar correlation were also found by Refs. \cite{Thi:2021jhz,Mondal:2022cva} using a semiagnostic EOS metamodel. Regarding the piecewise polytropes at higher densities, we only see the changes in first polytrope, $\Gamma_1$ first decreases with the addition of PREX, and then gradually increases with additional constraints. The only significant correlation among the EOS parameters we see, is between $L$ and $\Gamma_1$. The decrease of $\Gamma_1$ after adding PREX can be attributed to the constraints coming from tidal deformability. The information from PREX enters through $L$ only which tends to increase $L$. This should result in larger radii for stars with masses $\sim 1.4-1.7 M_\odot$. But, too large radii should be excluded due to the tidal deformability constraints from GW170817. Therefore, $\Gamma_1$ decreases to compensate the stiffness resulting from large $L$. The interplay between these two parameters is also seen in the pressure-vs-energy density plot in Figure \ref{fig:eos-post}. We observe that the inclusion of HESS makes the EOS softer at lower densities to produce smaller radii for the NS sequences by the decrease of the slope of symmetry energy. But, the requirement of $\sim 2M_{\odot}$ makes the EOS stiffer at high density thereby increasing the first polytrope. The absolute maximum masses from these EOSs stay unaffected because of no significant changes in $\Gamma_2$ and $\Gamma_3$. 

In Figure \ref{fig:mr-post}, we have shown the mass-radius posterior of the four cases corresponding to the EOS posteriors, shown in Figure \ref{fig:eos-post}. Here, we can clearly see the effect of the constraints on the structure of the NSs. The maximum mass regions for these constraints remain mostly unchanged following the same trend in the EOS posteriors. But, the radii of the low mass stars become larger after adding PREX, then gradually smaller with the addition of CREX and HESS. This is consistent with the decrease of the slope of symmetry energy at saturation in Figure \ref{fig:eos_params}. 

The correlations among the astrophysical observables ($R_{1.4}, \Lambda_{1.4}, M_{\rm max}$) and the nuclear parameters are shown in figure \ref{fig:macro_params}. Here, we show only $L$ and $K_{\rm sym}$ among the nuclear parameters as we have seen from figure \ref{fig:eos_params}, these are the ones most affected by the constraints. We also predict the central densities of the maximum mass stars from our analysis. With Astro, we get the maximum mass $\sim 2.17 M_\odot$ with the central pressure $\sim 181$ MeV fm$^{-3}$. Then, we add PREX to find a slight decrease in $M_{\rm max}$ to $\sim 2.15 \odot$. After adding CREX, it increases to $\sim 2.17 M_\odot$, indicating slight stiffness at that densities. As a result, the central pressure decreases slightly to $\sim 182$ MeV fm$^{-3}$. However, the radius of $1.4 M_\odot$ NS, $R_{1.4}$ behaves in an opposite way. We find that including CREX reduces $R_{1.4}$ from $\sim 12.75$ km to $\sim 12.66$ km, thereby indicating a softness around that densities. Although the change in the tidal deformabilities is negligible. Before adding HESS, the $\Lambda_{1.4}$ remains around $\sim 530$, consistent with the findings of Ref. \cite{Biswas:2021yge}. Finally, the addition of HESS decreases both $R_{1.4}$, $\Lambda_{1.4}$ and $M_{\rm max}$. While the decrease in $M_{\rm max}$ is very small, to $\sim 2.14 M_\odot$, the $R_{1.4}$ becomes $\sim 12.2$ km. The tidal deformability decreases to $\sim 447$. We can comment that the softening due to HESS occurs over the whole range of densities, small at very high densities, but significant at lower densities. This is again consistent from the behavior of $L$. In figure \ref{fig:macro_params}, we see a hint of correlation of $L$ with $R_{1.4}$, but not with $M_{\rm max}$. This type of correlation was also seen in previous studies \cite{Char:2023fue} where a relativistic hadronic model was used.

Now, we discuss the properties of HESS J1731-347 calculated within the hybrid+PP model. In figure \ref{fig:HESS_prop}, we have shown the inferred mass, radius, tidal deformability, and central pressure within $90\%$ CI. We infer its mass $0.83^{+0.26}_{-0.19} M_\odot$ and radius $11.78^{+0.81}_{-0.74}$ km. Since it is a relatively smaller object, its median of tidal deformability ($\sim 6340$) is on the higher side. We also estimate its central pressure to be $25.10^{+18.11}_{-9.09}$ MeV fm$^{-3}$. This corresponds to a slightly higher pressure than the value of pressure at saturation commonly considered \cite{Glendenning:1997wn}. In figure \ref{fig:HESS_MRs}, we compare the observed M-R data against the inferred values in this work. Although, we have found an overlap at $90\%$ CI, the radii do not overlap at $68\%$ CI. We also see a sharp bound on the lower bound on the inferred mass. This is a physical bound signifying that it is not possible to build a lower mass configuration beyond that region within our nucleonic EOS model that satisfies the constraint on the radius simultaneously. This is the limit of the hybrid+PP model on the low mass, low radius side. If we observe a new object even smaller than this object both in its size and mass, we have to invoke a different hypothesis regarding the composition of such object.

\section{Conclusion}
\label{summary}
In this work, we have presented a full statistical analysis of the impact of the mass-radius measurements of the central compact object of HESS J1731-347 on NS matter consisting of nucleons. We have used the full M-R posterior of HESS J1731-347 while performing the Bayesian analysis. We have also shown how the neutron skin thickness measurements affect NS observables. We have systematically studied the effect of the combinations of these constraints on the symmetry energy and its derivatives ($L$ and $K_{\rm sym}$). We have found the inferred EOS posterior becomes stiffer due to the effect of PREX, but becomes substantially softer with HESS. The softening is more prominent at lower densities than at higher densities as the maximum mass remains mostly unchanged. We have to keep in mind the systematic uncertainties in the modelling of X-ray spectra from HESS J1731-347, as mentioned in several works \cite{Alford:2023waw,Koehn:2024set}. Given such a star exists, we have found that this measurement can push our EOS model to its limit. 

While we were conducting this study, a new simultaneous mass-radius measurement of the nearest and the brightest millisecond pulsar PSR J0437-4715 was published by the NICER collaboration \cite{Choudhury:2024xbk}. The new data matches well with our results at the $90\%$ CI. Hence, we have not incorporated it into our analysis. However, we plan to conduct a thorough study to quantify the effect of the new data.

\section*{Acknowledgements}
This work has been partially supported by the Fonds de la Recherche Scientifique-FNRS, Belgium, under grant No. 4.4501.19. PC is supported by European Union's HORIZON MSCA-2022-PF-01-01 Programme under Grant Agreement No. 101109652, project ProMatEx-NS.  BB acknowledges the support from the Deutsche Forschungsgemeinschaft (DFG, German Research Foundation) under Germany's Excellence Strategy – EXC~2121 ``Quantum Universe'' –
390833306 and the Alexander von Humboldt Foundation through a Humboldt Research Fellowship for Postdoctoral Researchers. This project has received funding from the
European Union’s Horizon 2020 research and innovation
programme under the Marie Skłodowska-Curie Grant
Agreement No. 101034371

\bibliography{mybiblio}
\end{document}